\RequirePackage{fix-cm}
\documentclass[twocolumn]{svjour3}          
\smartqed  
\usepackage[super,sort&compress,comma]{natbib} 
\usepackage{mhchem}

\usepackage{graphicx} 

\usepackage{array}
\usepackage{color}
\usepackage{amssymb}
\newcommand{\un}[1]{\ensuremath{\unskip\,\mathrm{#1}}}
\newcommand{\vect}[1]{\mathbf{#1}}
\DeclareGraphicsExtensions{.pdf}

\begin{document}

\title{Why the aspect ratio? Shape equivalence for the extinction spectra of gold nanoparticles.}
\titlerunning{Shape equivalence for the extinction spectra of gold nanoparticles}
\author{Doru Constantin
}                     
%
%
\institute{D. Constantin \at Laboratoire de Physique des Solides, CNRS, Univ. Paris-Sud,\\ Universit\'{e} Paris-Saclay, 91405 Orsay Cedex, France\\
	\email{doru.constantin@u-psud.fr}}

\date{Received: date / Revised version: date}
%
\maketitle

\begin{abstract}
We compare the light extinction spectra of elongated gold nanoparticles with different shapes (cylinder, spherocylinder and ellipsoid) and sizes of 10 to 100~nm. We argue that the equivalence of the various moments of mass distribution is the natural comparison criterion --rather than the length-to-diameter (aspect) ratio generally used in the literature-- and that it leads to better spectral correspondence between the various shapes. 	
\keywords{gold nanorods \and UV-Vis-IR \and absorbance \and plasmon resonance \and aspect ratio \and inertia tensor}
\PACS{46.05.+b \and 78.67.-n \and 78.67.Qa}

\end{abstract}

\section{Introduction}\label{sec:Intro}

Over the last two decades, the striking optical properties of noble metal nanoparticles have raised considerable interest. Among the variety of morphologies obtainable by chemical synthesis, ``nanorods'' (elongated cylindrical or, more commonly, spherocylindrical particles) are some of the most studied, due to their strong longitudinal resonance, which can be tuned by varying their anisotropy. 

Silver and gold nanorods were thoroughly studied by UV-Vis-IR spectroscopy. Although numerical simulation techniques are nowadays quite accessible, see e.g. Refs. \citenum{Jain:2006,Payne:2006,Khlebtsov:2007b,Khlebtsov:2011,Gonzalez:2013}, these particles are often modeled as ellipsoids \cite{Skillman:1968,vanderZande:2000,Brioude:2005,Chen:2013}, since this shape has the great advantage of being amenable to analytical treatment, as briefly discussed in the following; see \cite{Quinten:2010} for a detailed and up-to-date discussion. The scattering problem was solved exactly for a sphere \cite{Mie:1908} and for an arbitrary spheroid \cite{Asano:1975}, but the expressions are cumbersome. In the Rayleigh (or electrostatic) limit, analytical approximations can be derived for ellipsoids \cite{Gans:1912}. They are valid for particles much smaller than the wavelength, a condition often fulfilled in practice. More elaborate but still tractable approximations exist, for spheres \cite{Meier:1983} and spheroids \cite{Wokaun:1985,Kelly:2003,Kuwata:2003,Moroz:2009} beyond the Rayleigh regime.

Unfortunately, the currently prevailing opinion  \cite{Lee:2005,Prescott:2006,Nehl:2008,Myroshnychenko:2008} is that the ellipsoid model does not accurately describe more realistic shapes, in particular with respect to the longitudinal plasmon resonance (that associated to the longest dimension). The authors reach this conclusion by comparing the position of the longitudinal plasmon peak (LPP) for cylinders, spherocylinders, and prolate spheroids with the same length $L$ and transverse diameter $D$ and hence the same aspect ratio $R=L/D$. \cite{Prescott:2006}. However, the literature provides no justification for using $R$ as the comparison parameter.

We consider the relevant criterion for identifying the ellipsoid corresponding to a given particle. In contrast with the established procedure discussed above, we argue that the relevant quantities are the various moments of the mass distribution, leading to an effective aspect ratio $R_{\text{eff}}$. The spectra of particles with different shapes: cylinder, spherocylinder and ellipsoid and the same $R_{\text{eff}}$ agree much better than those for an equal aspect ratio $R$.

\section{Model}\label{sec:Meth}

The most general description of a mass distribution, for a body or a system of particles), is in terms of its various moments \cite{Heard:2006}. The first of these are, in increasing tensorial order, $m$ (the total mass of the system), $m \, \vect{r_{\text{CM}}}$ (with $\vect{r_{\text{CM}}}$ the center-of-mass vector) and the inertia tensor:
\begin{equation}
\tens{I} = \int_{V} \mathrm{d} \vect{r} \rho(\vect{r}) \left[ (\vect{r} - \vect{r_{\text{CM}}})^2 \tens{1} -  (\vect{r} - \vect{r_{\text{CM}}}) \otimes  (\vect{r} - \vect{r_{\text{CM}}})\right]
\end{equation}
\noindent where $V$ is the volume of the body, $\rho(\vect{r})$ is the local mass density, $\tens{1}$ is the unit tensor and $\otimes$ denotes the outer product. $\tens{I}$ is widely used in mechanics, but is also relevant for the interaction of particles with radiation: for instance, the gyration radius $R_g$ defined as $m R_g^2 = \frac{1}{2} \text{Tr} \tens{I}$ is extensively used in small-angle scattering techniques \cite{Guinier:1955}. This strategy is also very similar to the traditional way of describing a charge distribution by its total charge, dipole, quadrupole and higher multipolar moments.

In this framework, the ellipsoid equivalent to a given body is the one with the same distribution moments. Geometrically, an ellipsoid is completely defined by three parameters (e.g. the semi-axes) chosen such that the three eigenvalues of $\tens{I}$ are the same as those of the initial body. The particle position $\vect{r_{\text{CM}}}$ does not influence the absorbance; as to the mass, we will return to it in Appendix~\ref{sec:appMass}.

\begin{figure}
\includegraphics[width=0.45\textwidth,angle=0]{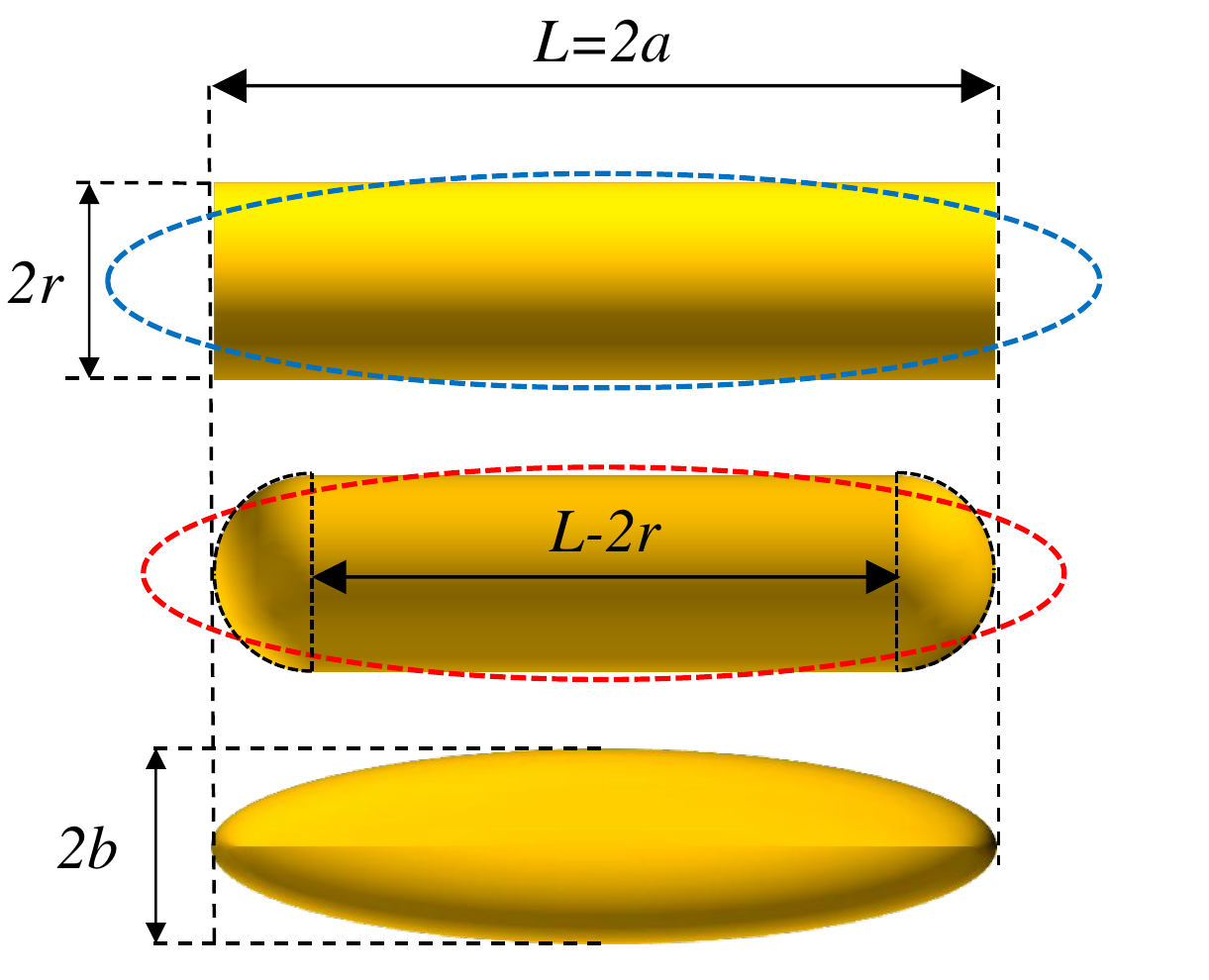}%
\caption{\label{fig:geom}Three nanoparticles (from top to bottom: cylinder, spherocylinder and prolate ellipsoid) with the same total length $L = 2a$ and diameter $D = 2r = 2b$, and thus the same aspect ratio $R = L/D = 5$. For the top two shapes, the superimposed contour represents the equivalent ellipsoid (see text).}%
\end{figure}
We studied three types of particles: cylinders, spherocylinders and prolate ellipsoids, described by their aspect ratio $R=\frac{L}{2 r}$ (exemplified in Figure~\ref{fig:geom} for $R=5$.) and considered homogeneous: $\rho(\vect{r}) = \rho$. For ellipsoids, one can also write $R = a/b$ (the ratio of the two semi-axes). All these shapes have azimuthal symmetry around their length, taken as parallel to $z$, so that only $I_x=I_y = \rho \int_{V} \mathrm{d} \vect{r} \left( y^2 + z^2 \right) $ and $I_z = \rho \int_{V} \mathrm{d} \vect{r} \left( x^2 + y^2 \right)$ need to be determined.
\begin{subequations}
\begin{equation}
\left\lbrace
\begin{array}{ll}
I_x &= \dfrac{m_{\text{cyl}}}{12} (3 r^2 + L^2)\\
I_z &= \dfrac{m_{\text{cyl}}}{2}  r^2
\end{array}
\right. \qquad \text{cylinder}\\
\label{eq:Icyl}
\end{equation}
\begin{equation}
\left\lbrace
\begin{array}{ll}
I_x &=\rho \pi r^5 \left \lbrace \dfrac{R-1}{6} \left [ 3 + 4(R-1)^2\right ] \right. \\
 &+ \left. \dfrac{4}{3} \left [ \dfrac{83}{320} + \left ( R - 1 + \dfrac{3}{8}\right )^2 \right ] \right \rbrace\\
I_z &=\rho \pi r^5 \left [ (R-1) + \dfrac{8}{15}\right ]
\end{array}
\right.\, \text{spherocylinder}\\
\label{eq:Isphcyl}
\end{equation}
\begin{equation}
\left\lbrace
\begin{array}{ll}
I_x &= \dfrac{m_{\text{ell}}}{5} (a^2 + b^2)\\
I_z &= \dfrac{2m_{\text{ell}}}{5}  b^2
\end{array}
\right.\qquad \text{ellipsoid}
\label{eq:Iellips}
\end{equation}
\label{eq:Iall}
\end{subequations}
\noindent where $m_{\text{cyl}}$ and $m_{\text{ell}}$ are the masses of the cylinder and ellipsoid, respectively. For the spherocylinder, the moments of inertia are not easily expressed as a function of the total mass of the object and are therefore given in terms of the geometrical parameters and of the mass density $\rho$.

For a given cylinder or spherocylinder we determine $I_x$ and $I_z$ as a function of $r$ and $R$ from \eqref{eq:Icyl} or \eqref{eq:Isphcyl} and identify them with \eqref{eq:Iellips}, yielding the semi-axes of the equivalent ellipsoid $a$ and $b$ or, equivalently, $b(r,R)$ and $a/b = R_{\text{eff}} (R)$. It is easily checked that $R_{\text{eff}}$ does not depend on the radius $r$:
\begin{equation}
\begin{array}{lll}
R_{\text{eff}}^{\text{cyl}} (R)&= \dfrac{2}{\sqrt{3}} R &\text{cylinder} \\
R_{\text{eff}}^{\text{sc}} (R)&= \sqrt{1 + \dfrac{4 \gamma }{3} \, \dfrac{\gamma ^2 + 2 \gamma + 3/4}{\gamma + 8/15}} &\text{spherocylinder}\\
R_{\text{eff}}^{\text{ell}} (R)&\equiv R &\text{ellipsoid}
\end{array}
\label{eq:Reff}
\end{equation}
\noindent where $\gamma = R-1$.

We performed numerical simulations of the extinction cross-section $Q_{\text{ext}}$ of anisotropic gold nanoparticles using the discrete dipole approximation (DDA) code \textsc{DDSCAT 7.3} \cite{Draine:1994,Flatau:2012} with the filtered coupled dipole method \cite{Yurkin:2010}. The refractive index of bulk gold $n_{\text{Au}}$ is that given by Johnson \& Christy \cite{Johnson:1972}, with no corrections for boundary dissipation. The ambient medium is water, described by a constant refractive index $n_{\text{H$_2$O}} = 1.33$. The particles are discretized using 90 dipoles along the diameter and $90 \cdot R$ along the length, where $R$ goes from 1 to 8. The particle radius is 5~nm\footnote{For simplicity, we compare particles with the same radius. Note, however, that identifying the moments in \eqref{eq:Iall} leads to slightly different radii for the ellipsoids as compared to the target particles. We checked that the resulting difference in peak position is negligible.}.
The electric field is parallel to the long axis of the particles (along $z$) in order to probe the LPP.

In the transverse configuration (electric field perpendicular to $z$), the spectrum changes very little with the aspect ratio \cite{Noguez:2007}. For randomly oriented particles, e.g. in colloidal solution, the overall spectrum is a superposition of the longitudinal and transverse components, so that a satisfactory description of the longitudinal spectrum ensures that the total spectrum is also correctly described.

To extend the results to larger particle sizes (10 and 20 nm radii) we performed additional simulations using the boundary element method (BEM) \cite{deAbajo:2002} as implemented by the \textsc{MATLAB} package \textsc{MNPBEM} \cite{Hohenester:2012}. The two methods yield very similar results, as shown in Appendix~\ref{sec:DDABEM}.

\section{Results}\label{sec:Res}

\begin{figure*}
\includegraphics[width=0.9\textwidth,angle=0]{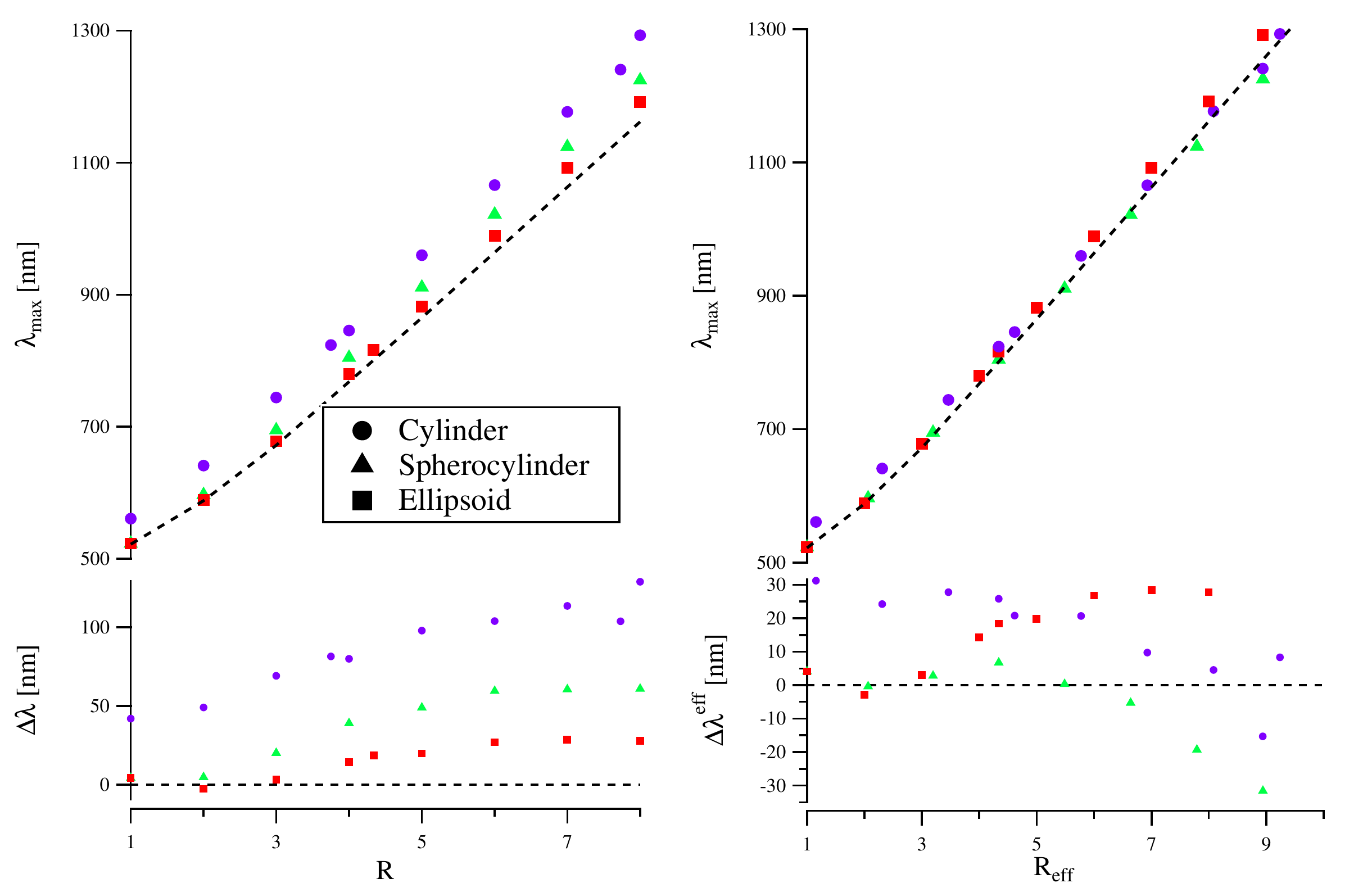}%
\caption{\label{fig:simul}Top: position of the LPP $\lambda _{\text{max}}$ in the DDA simulations for cylinders $(\bullet)$, spherocylinders $(\blacktriangle)$ and ellipsoids $(\blacksquare)$ and for ellipsoids in the Rayleigh limit (dashed line). Bottom: Differences $\Delta\lambda$ and $\Delta\lambda ^{\text{eff}}$ between the simulation results and the Rayleigh limit (note the difference in scale range between these two parameters). Left: Values plotted as a function of the aspect ratio $R$. Right: Values plotted as a function of the effective aspect ratio $R _{\text{eff}}$. The radius ($r$ or $b$) is 5~nm for all particles. The top left panel can be directly compared to Figure~3a of Prescott \& Mulvaney \cite{Prescott:2006}.
}%
\end{figure*}

\nocite{Bohren:1983}
We fitted the extinction spectrum with the sum of a Lorentzian peak and a cubic background to determine the position of the LPP, plotted in Figure~\ref{fig:simul} for the three shapes, as a function of $R$ (left panel) and $R _{\text{eff}}$ (right panel). As reference, we also added the LPP position for ellipsoids in the Rayleigh limit, using the Gans formula (Ref. \citenum{Bohren:1983}, Eq. (5.32)).

As $R$ increases, so does the difference in LPP position between different shapes. When plotted against $R _{\text{eff}}$, however, the LPP values are much closer together and the variation is non-monotonic.

\begin{figure*}
	\includegraphics[width=0.9\textwidth,angle=0]{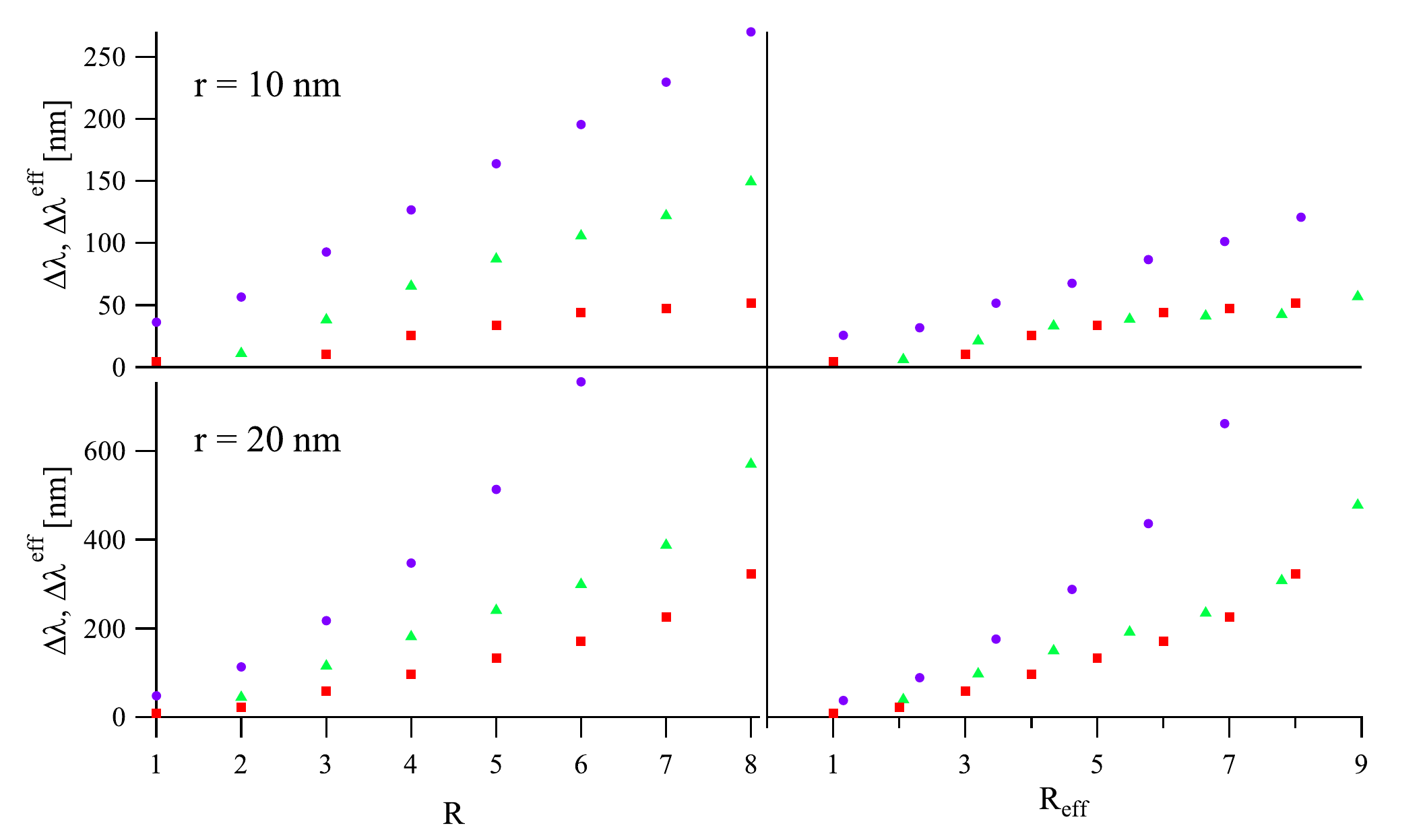}%
	\caption{\label{fig:r10r20} Differences between the simulation results and the Rayleigh limit $\Delta\lambda$ as a function of the aspect ratio $R$ (left) and $\Delta\lambda ^{\text{eff}}$ as a function of the effective aspect ratio $R _{\text{eff}}$ (right) for particles with $r=10 \un{nm}$ (top) and with $r=20 \un{nm}$ (bottom). Symbols are as in Figure~\ref{fig:simul}.
	}%
\end{figure*}

The tendency also holds for larger particles, as shown in Figure~\ref{fig:r10r20} for radii of 10 and 20~nm. For clarity, only the differences with respect to the Rayleigh limit are shown, corresponding to the bottom of Figure~\ref{fig:simul}. The peak positions for the spherocylinder and ellipsoid are remarkably close together, while those of the cylinder exhibit a red shift which is increasingly pronounced with the aspect ratio and the radius. This is probably due to field concentration at the sharp edges of the cylindrical particles, a localized effect that cannot be captured by our simplified model.

\section{Comparison with experimental data}\label{sec:Exp}

\begin{figure}
	\includegraphics[width=0.45\textwidth,angle=0]{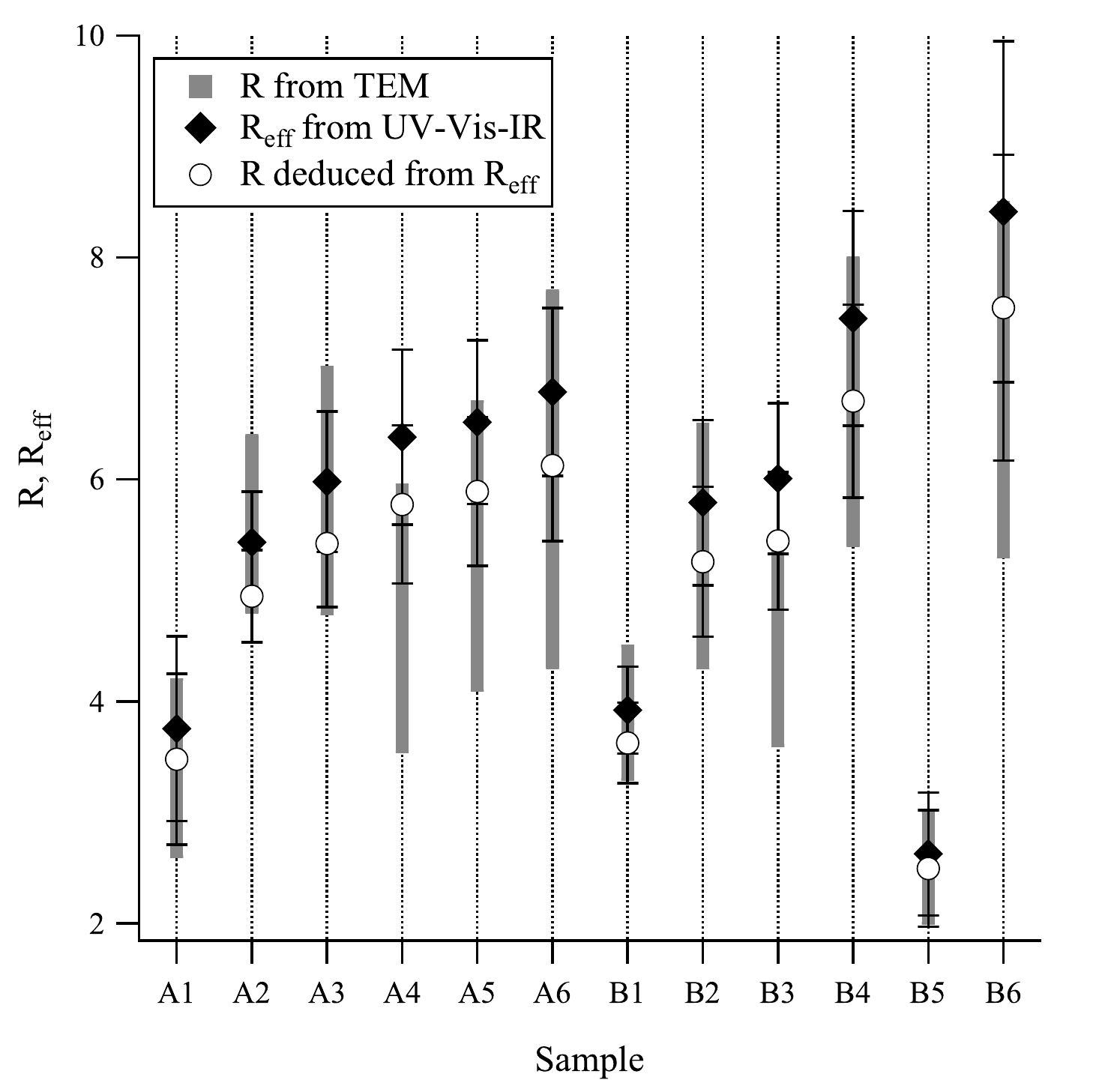}%
	\caption{Aspect ratio $R$ determined by transmission electron microscopy (TEM) (grey bars) and effective aspect ratio $R_{\text{eff}}$ measured by UV-Vis-IR spectroscopy (solid diamonds). The geometric aspect ratio for spherocylinders $R$ (open dots) is then extracted from $R_{\text{eff}}$ by inverting Eq.~\eqref{eq:Reff}. The error bars (UV-Vis-IR) and bar height (TEM) indicate the mean $\pm$ standard deviation: $R \pm \sigma _R$ or $R_{\text{eff}} \pm \sigma_{\text{Reff}}$. Based on Figure~5 of Ref.~\citenum{Slyusarenko:2014}.\label{fig:AR}}%
\end{figure}

We also checked our model against the experimental data measured for gold nanorods with varying aspect ratio in Ref.~\citenum{Slyusarenko:2014}. The mean aspect ratio $R$ and its standard deviation $\sigma _R$ are determined via TEM and are shown in Figure~\ref{fig:AR} as grey bars versus the sample code. The UV-Vis-IR spectroscopy curves are fitted with an ellipsoid model, yielding the distribution of \emph{effective} aspect ratio. Its mean $R_{\text{eff}}$ and standard deviation $\sigma_{\text{Reff}}$ are shown as solid diamonds with error bars.

If our model is correct, the $R$ values (open dots) obtained from the experimental $R_{\text{eff}}$ distribution by inverting the middle relation in Eq.~\eqref{eq:Reff} are the true geometrical ones and should coincide with the TEM results. For most samples this is indeed the case, confirming the improvement. One should however keep in mind that TEM only samples a very small fraction of the particles (those deposited on grids and, among these, only those visible in the images) in contrast with  UV-Vis-IR spectroscopy, which averages over all particles contained in a few milliliters of solution. The latter technique should therefore be much more representative of the complete particle distribution.

\section{Discussion and Conclusion}\label{sec:Conc}

For ellipsoids in the Rayleigh approximation, the LPP is completely described by a geometrical factor $P_z$. One can then define an equivalent ellipsoid of a particle by fitting its simulated spectrum with the Mie-Gans formula and finding the corresponding $P_z$ \cite{Prescott:2006}. Our approach here is completely different: we find the equivalent ellipsoid of the particle by applying very general principles to its mass distribution and without any consideration of its electrostatic or optical properties.

We conclude that the LPP positions of particles with the same moments of inertia (defined in \eqref{eq:Iall}) are significantly closer than for particles with the same aspect ratio $R$.

In conclusion, when the mass distribution is properly taken into account, the effective ellipsoid approximation describes fairly well the optical properties of elongated gold nanoparticles, in particular those of spherocylinders. We therefore expect the results can be generalized to other rounded particles (e.g. dumbbells \cite{Song:2005}) but probably not to those exhibiting edges or tips (e.g. prisms \cite{Millstone:2009}), where the field concentration is significant. It would also be interesting to study the influence of faceting, observed in certain nanorods\cite{Wang:1999,Goris:2012}, on their optical spectra.

Combining this correction with an approximate analytical relation for the response of ellipsoids \cite{Moroz:2009} and accounting for the polydispersity in aspect ratio \cite{Eustis:2006,Amendola:2009,Slyusarenko:2014} should yield a quantitatively accurate model for the experimental extinction spectra.

\appendix

\section{Volume correspondence}\label{sec:appMass}

\begin{figure}
\includegraphics[width=0.45\textwidth,angle=0]{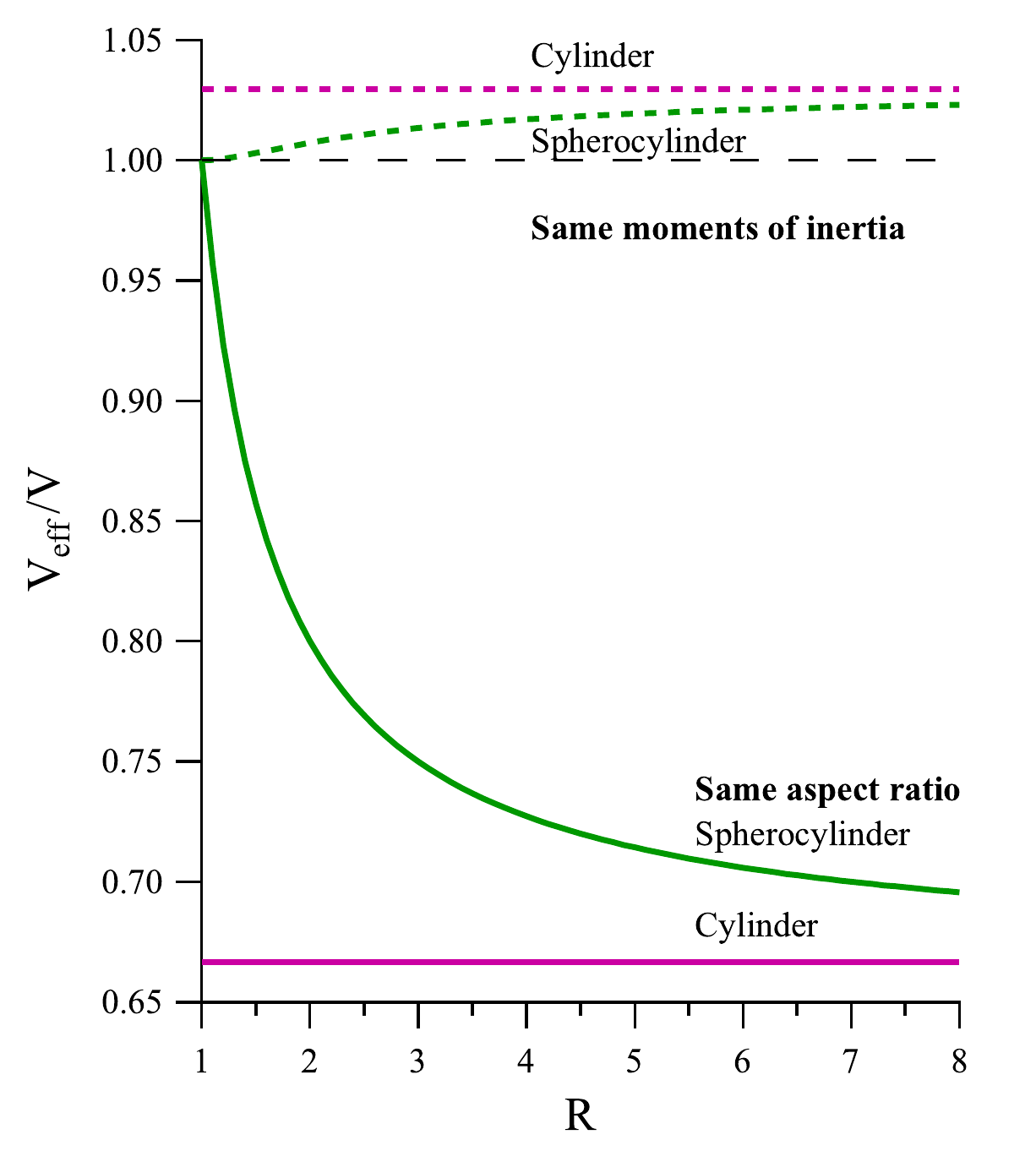}%
\caption{\label{fig:volume}Ratio of the volume $V_{\text{eff}}$ of the equivalent ellipsoid to the volume $V$ of the target particle (cylinder or spherocylinder), based on identifying the moments of inertia (dotted line) or by using the same aspect ratio (solid line), compared to the exact value of 1 (dashed line).}%
\end{figure}

The three moments of inertia $I_x, I_y$ and $I_z$ completely define the equivalent ellipsoid for a given particle, including its volume $V_{\text{eff}}$ and hence its mass $m_{\text{eff}} = \rho V_{\text{eff}}$. In Figure~\ref{fig:volume} we show the ratio of $V_{\text{eff}}$ to the volume $V$ of the target particle, for the cylinder and spherocylinder and for aspect ratios from 1 to 8. For reference, we also show the same ratio for ellipsoids having the same aspect ratio $R$ as the target particle.

\noindent Clearly, the volumes are much closer when the corresponding ellipsoid is chosen based on the moments of inertia (the discrepancy is below 3~\%) than based on the aspect ratio (where the discrepancy can reach 33~\%). This result is noteworthy on two counts: first, because it comforts our choice of the moments of inertia as relevant parameters and second because the volume must be correctly described for practical applications, e.g. when estimating the particle concentration in solution from the extinction spectrum.

\section{Agreement between the DDA and BEM methods}\label{sec:DDABEM}

\begin{figure*}
\includegraphics[width=0.45\textwidth,angle=0]{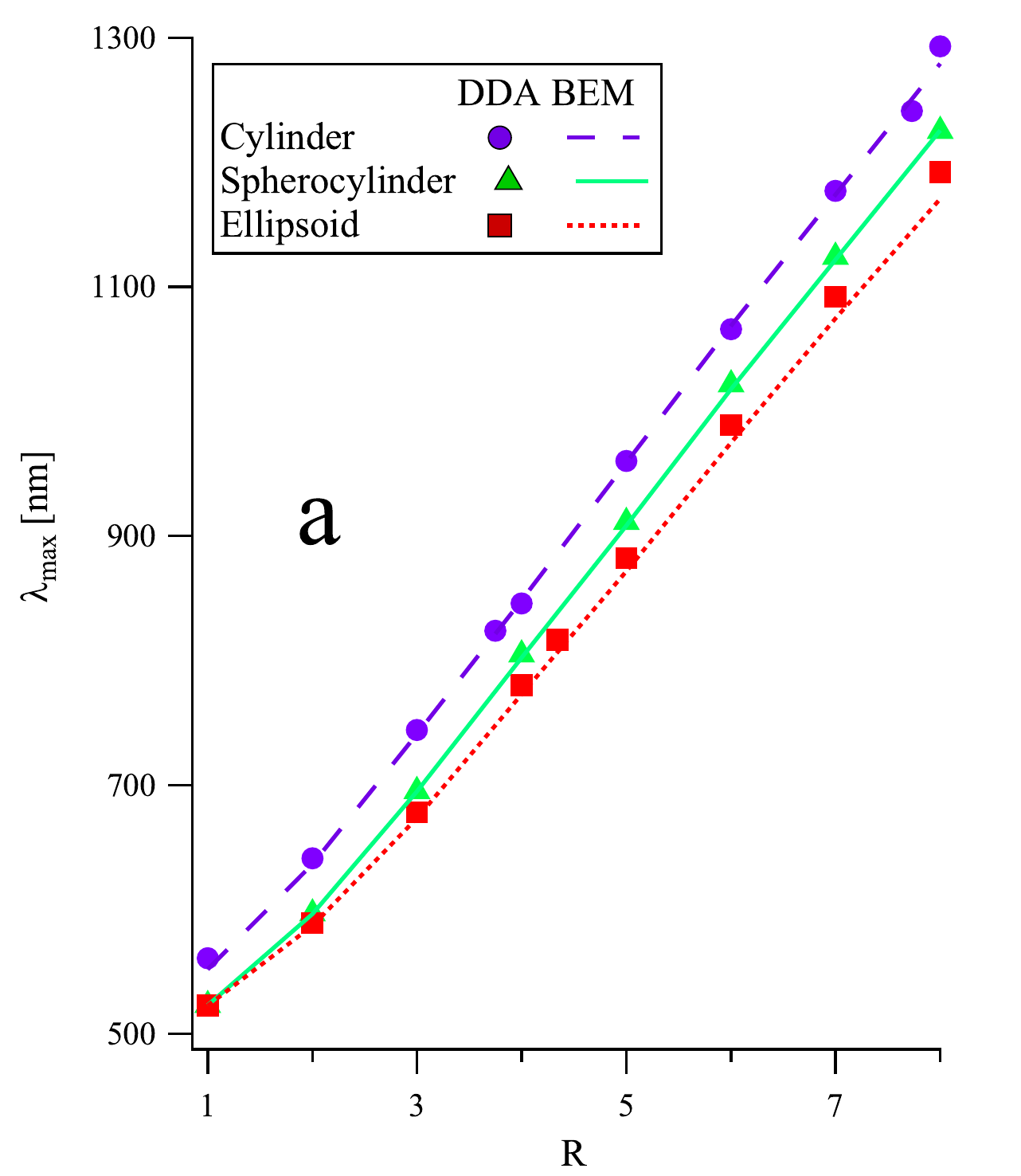}\includegraphics[width=0.45\textwidth,angle=0]{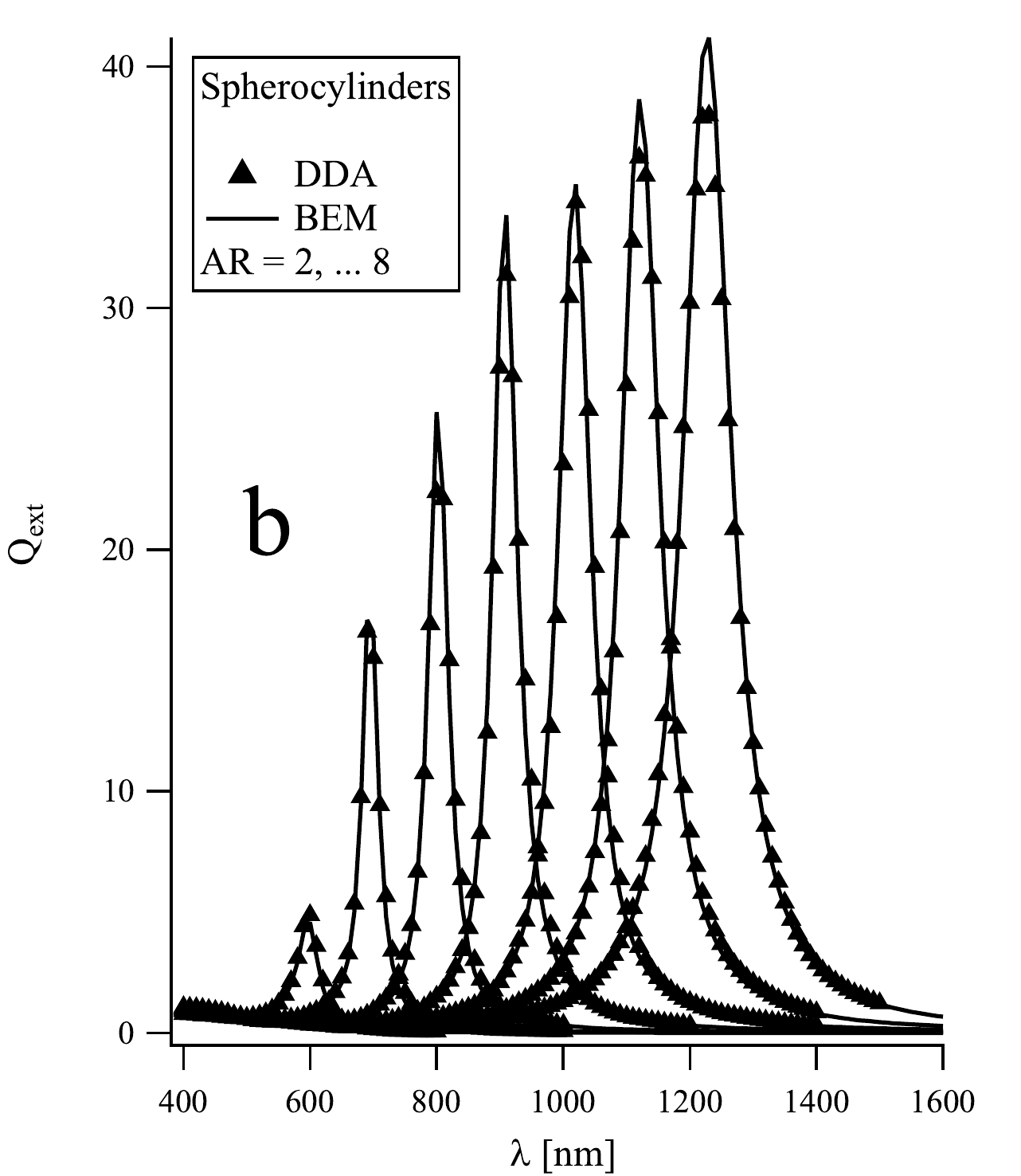}%
\caption{\label{fig:DDABEM}Comparison between the DDA and BEM methods. a) LPP position for the three particle shapes, with aspect ratios between 1 and 8, obtained via DDA (symbols) and BEM (lines). b) Comparison between the extinction spectra of spherocylinders, obtained via DDA (symbols) and BEM (lines), for aspect ratios between 2 and 8.}%
\end{figure*}

The DDA method is more time-consuming, so we only used it for the smaller particles (5~nm in radius) and employed the BEM technique for the larger objects. To make sure that the results are compatible, we also ran the BEM simulations for the small particles, with the results shown in Figure~\ref{fig:DDABEM}.
\noindent The agreement is excellent for the peak position (Figure~\ref{fig:DDABEM}a) as for the longitudinal spectrum (Figure~\ref{fig:DDABEM}b).

\begin{acknowledgements}
Kostyantyn Slyusarenko is acknowledged for fruitful discussions.
\end{acknowledgements}

\footnotesize{
	\bibliography{corresp} 
	\bibliographystyle{spphys} 
}


\end{document}